\documentclass[pra,twocolumn,superscriptaddress,preprintnumbers,amsmath,amssymb,showpacs,showkeys]{revtex4}
\usepackage{graphicx}
\usepackage{dcolumn}
\usepackage{bm}
\usepackage{epsfig}
\usepackage[dvipsnames]{color}
\usepackage{epstopdf} 

\newcommand{\beq}{\begin{eqnarray}}
\newcommand{\eeq}{\end{eqnarray}}

\newcommand{\rmd}{{\text d}}

\newcommand{\figurewidth}{\columnwidth}
\input xy 
\xyoption{all}
\definecolor{myback}{rgb}{1,.964,.8}

\begin{document}

\title{Continuous-Time Quantum Algorithms for Unstructured Problems}
\author{Itay Hen}
\email{itay.hen@nasa.gov}
\affiliation{Quantum Laboratory, Applied Physics Center, NASA Ames Research Center, Moffett Field, California 94035, USA}
\affiliation{Department of Physics, University of California, Santa Cruz, California 95064, USA}

\date{\today}

\begin{abstract}
We consider a family of unstructured problems, for which we propose a method for constructing analog, continuous-time quantum algorithms
that are more efficient than their classical counterparts. 
In this family of problems, which we refer to as `scrambled output' problems, one has to find a minimum-cost configuration of a given integer-valued $n$-bit function whose output values have been scrambled in some arbitrary way. Special cases within this set of problems are Grover's search problem of finding a marked item in an unstructured database, certain random energy models, and the functions of the Deutsch-Josza problem. 
We consider a couple of examples in detail. In the first, we provide a deterministic analog quantum algorithm to solve the 
seminal problem of Deutsch and Josza, in which one has to determine 
whether an $n$-bit boolean function is constant (gives $0$ on all inputs or $1$ on all inputs) or balanced (returns $0$ on half the input states and $1$ on the other half). We also study one variant of 
the random energy model, and show that, as one might expect, its minimum energy configuration can be found quadratically 
faster with a quantum adiabatic algorithm than with classical algorithms.
\end{abstract}

\maketitle

\section{\label{intro}Introduction}

The paradigm of Adiabatic Quantum Computation (AQC) proposed by Farhi {\it et al.}~\cite{farhi_long:01} about a decade ago is a simple yet intriguing approach to problem solving on a quantum computer. 
Unlike the leading paradigm of circuit-based quantum computing, AQC is an analog continuous-time method that does not require the design and use of quantum gates.
As such, it can in many ways be thought of as a simpler and perhaps more profound method for performing quantum computations 
that is also easier to implement experimentally~\cite{vandersypen:01,gaitan:12}.

Even though AQC has been shown to be polynomially-equivalent to circuit-based computation~\cite{aharonov:07,mizel:07},
and despite intensive research in the area (see, e.g., Refs.~\cite{hogg:03,farhi:02,farhi:08,altshuler:09,knysh:11,young:08,young:10,hen:11,hen:12} and references therein), to date, there are almost no clear-cut concrete examples for efficient quantum-adiabatic algorithms that reveal the potentially-powerful ``fully-quantum'' capabilities encompassed in AQC.
One possible reason for that is, presumably, that there is usually no obvious way to `tailor' the 
adiabatic algorithm to the specific problem being examined, and to make use of the structure of the problem to, for example, 
modify the beginning Hamiltonian in a clever way that would speed up the computation (a notable exception is Ref.~\cite{roland:03}). 
For most of the interesting optimization problems, being able to do so, may be as
hard as solving the original problem itself~\footnote{Exceptions to the rule are adaptive methods which attempted by repeatedly running the QAA, performing measurements at the end, to adjust to some extent the driver Hamiltonian
for in order to obtain some speedup in the calculation~\cite{dickson:11}.}. 

Interestingly, one of the few problems for which quantum speedup has been obtained in the context of AQC, is 
Grover's unstructured search problem~\cite{grover:97}, in which one searches for a marked item in an unstructured database.
Roland and Cerf~\cite{roland:02} have demonstrated that while the application of the
adiabatic algorithm to Grover's problem with a linear rate results in a running time that is of order $N$ -- $N$ being the number of items in the database -- a carefully chosen variable rate of the adiabatic parameter yields a running time that scales like $\sqrt{N}$, i.e., a quadratic speed-up is gained, similarly to the original result by Grover found for the circuit-based model~\footnote{We note that a continuous-time quantum algorithm 
for Grover's search problem which exhibited a quadratic speedup was initially proposed by Farhi and Gutmann, albeit as a non-adiabatic quantum computation~\cite{farhi:98}.}.  

The adiabatic algorithm for Grover's problem utilizes the concept of local adiabatic evolution,
in the framework of which the adiabatic parameter is varied not at a constant rate 
but rather at a variable rate, slowing down in the vicinity of the minimum gap and speeding-up in places where the gap is large.

Local adiabatic evolution however can only be efficiently used in cases where one has proper knowledge of the exact behavior of the gap and relevant matrix elements of the system for the problem in question. This is normally not the case.
In the Grover problem, the ability to compute the gap and matrix element of the problem stems from prior knowledge of the spectrum of the problem Hamiltonian, which, ultimately, reduces the problem into a simple two-level system~\cite{roland:02}.

In what follows, we consider a family of unstructured problems, which we refer to as `scrambled output' models,
and show how one may utilize knowledge of the spectrum of the Hamiltonian of the problem to find analog, continuous-time, 
algorithms that are more efficient than their classical analogues.  In that sense, this family of problems is a generalization, or an extension, 
to the problem solved by Roland and Cerf. 

In scrambled output problems, one has to find a minimum input configuration (i.e., $\arg \min$) of an $n$-bit function whose set of outputs (and their multiplicities) is given in advance, up to an unknown constant offset. The exact mapping between the $N=2^n$ input configurations and the various outputs is also not given (i.e., it is as though the outputs of a known function have been scrambled in some arbitrary way). 
As we also discuss later, special cases in this family of problems are the unstructured database search problem considered by Roland and Cerf~\cite{roland:02}, certain variants of the random energy model and the functions of the Deutsch-Josza problem. 

We illustrate the manner in which AQC may be used to solve scrambled output problems by constructing 
efficient analog continuous-time algorithms for two specific examples: the Deutsch-Josza~\cite{deutsch:92} problem for which we find an 
efficient $O(1)$ deterministic solution, and a variant of the random energy model, for which the minimum energy configuration is found quadratically faster than the corresponding classical algorithms.

The paper is organized as follows. In the next section we briefly discuss the principles of the Quantum Adiabatic Algorithm 
that is the heart of AQC, and with which the above models are solved. In Sec.~\ref{sec:sop}, we describe scrambled output problems  in detail. We then study two examples. In Sec~\ref{sec:dj}, we suggest an analog algorithm for the Deutsch-Josza problem,
and in Sec.~\ref{sec:rem} we consider a variant 
of the random energy model. Finally, we conclude with a few comments in Sec.~\ref{sec:conc}. 

\section{\label{sec:qaa}Quantum Adiabatic Algorithm (QAA)}

The Quantum Adiabatic Algorithm (QAA) provides the general approach for solving optimization problems 
on an analog continuous-time quantum computer~\cite{farhi_long:01}.  
Within the framework of the QAA,
the solution to an optimization problem is encoded in the ground state of a Hamiltonian
$\hat{H}_p$. 
To find the solution, the QAA prescribes the following course of action. As a first
step, the system is prepared in the ground state of another `driver' Hamiltonian
$\hat{H}_d$.  The driver
Hamiltonian is chosen such that it does not commute with the problem
Hamiltonian and has a ground state that is fairly easy to prepare. 
As a next step,
the Hamiltonian of the system is slowly modified from $\hat{H}_d$ to
$\hat{H}_p$, using the linear interpolation, i.e.,
\begin{equation}
\hat{H}(s)=s \hat{H}_p +(1-s) \hat{H}_d \,,
\end{equation}
where $s(t)$ is a parameter varying smoothly with time,
from $0$ at $t=0$ to $1$ at the end of the algorithm,
$t=\mathcal{T}$.  If this process is done slowly enough, the
adiabatic theorem of Quantum
Mechanics (see, e.g., Refs.~\cite{kato:51} and~\cite{messiah:62})
ensures that the system will stay close to the ground state of the
instantaneous Hamiltonian throughout the evolution, so that one finally
obtains a state close to the ground state of $\hat{H}_p$.  At this point,
measuring the state will give the solution of the original problem with high
probability. 

For adiabatic processes, the adiabatic profile function $s(t)$ must be chosen such that the evolution of the system is slow. 
In the simple case where $s(t)$ varies from zero to one at a constant rate,  
the runtime $\mathcal{T}$ must be chosen to be large enough so that the adiabatic
approximation holds: this condition determines the
efficiency, or complexity, of the QAA.  A condition on $\mathcal{T}$ can
be given in terms of the instantaneous eigenstates $\{ | m
\rangle \}$ and eigenvalues $\{E_m \}$ of the Hamiltonian $H(s)$,
as~\cite{wannier:65,farhi:02}
\begin{equation} \label{eq:T}
\mathcal{T} =\frac1{\epsilon} \, {\max_{s} V_{01}(s)  \over
\min_s g^2(s)} \,,
\end{equation} 
where $g(s)$ is the first
excitation gap $E_1(s)-E_0(s)$ and $V_{01}(s) = \left| \langle 0 | \rmd H / \rmd s | 1\rangle \right|$ (in our units $\hbar=1$).
Here, $\epsilon$ is a small number inversely proportional to the running time of the algorithm. 
The smaller $\epsilon$ is chosen to be, the slower the evolution of the system will be and the larger the probability of success will become.
For an adiabatic process to have zero error, the running time must tend to infinity.

As discussed in the Introduction, one could gain significant speed-up in calculation times 
in cases where the concept of local adiabatic evolution~\cite{roland:02} can be utilized.
This is done by formulating a `local' Landau-Zener condition for each value of the adiabatic 
profile function $s(t)$, namely:
\beq \label{eq:lae}
\left| \frac{\rmd s }{\rmd t} \right| \leq \epsilon  \frac{g^2(s)}{V_{01}(s)} \,,
\eeq
where $\epsilon$, $g(s)$ and $V_{01}(s)$ are as in Eq.~(\ref{eq:T}). 

Before moving on to describe scrambled output problems, we note here that
in special cases, the `machinery' of the QAA may also be used differently than described above, namely, in a not-necessarily adiabatic fashion. 
By `non-adiabatic' use, it is  meant  that the state of the system may `wander away', at least to some extent, from the 
instantaneous ground-state of the system in the course of the algorithm before returning to it towards the end
of the run. This concept is discussed later in greater detail. One known example for such `non-adiabatic' computation
is the analog version of Grover's search algorithm suggested by Farhi and Gutmann in Ref.~\cite{farhi-2008-6} . 

\section{\label{sec:sop}Scrambled output problems}

In what follows, we consider a family of problems, that we refer to as `scrambled output' problems, and which, as we illustrate next, 
can be solved more efficiently on an analog quantum computer than on a classical computer. As we explain in what follows, this 
family of problems is a generalization, or an extension, of the unstructured database search considered by Roland and Cerf~\cite{roland:02}. 

In `scrambled output' problems, one is asked to find a minimizing input 
configuration of an integer-valued $n$-bit function whose $K+1$ output values $f_0<f_1<\ldots<f_K$ and their multiplicities
$m_0,m_1,\ldots,m_K$ are known in advance, up to an unknown constant offset, denoted by $e_0$, that is added to all outputs. 
Here, we shall assume that the number of distinct output values, $K+1$ is much smaller than the $2^n$ actual eigenvalues.  
While the output values of the function are known, it is not known which output belongs to which input bit-configuration; the output values are `scrambled' in some arbitrary way. The additional unknown constant offset serves to further complicate the problem: It makes the minimum-energy configuration harder to find for classical algorithms, but, as we shall see, will have no effect on the quantum ones.

Special cases in this family of problems are: i) The original Grover's search problem considered by Roland and Cerf~\cite{roland:02} in which $f_0=0$ with multiplicity $m_0=M$ (where $M$ is the number of marked items or `targets') and $f_1=1$ with $m_1=N-M$. 
ii) The Deutsch-Josza input-functions
(which we discuss in the next section) for which both the balanced and constant functions are described by $f_0=0$ and $f_1=1$, with multiplicities $m_0=m_1=N/2$ (balanced) and $m_{0/1}=0$ and $m_{1/0}=N$ (constant). iii) A variant of the `random energy model' for which $f_j=j$ with $j=0..n$ and $m_j=\binom{n}{j}$ ($n$-choose-$j$). This model has been previously considered in Ref.~\cite{farhi:11} in the context of the QAA. 

Since the output values of the given function are scrambled in an arbitrary way, 
the typical running time  of any classical search algorithm designed to find a minimum-cost configuration will scale as $O(N)$ (unless of course there are exponentially many minimizing configurations). For example, a classical deterministic solution to the problem will require in the worst case $N-m_0$ evaluations of the scrambled-output function. 

In quantum-algorithmic terminology, the problem described above can be encoded in a matrix $F$ (which we take to be diagonal
in the computational basis),
whose eigenvalues are known but have been shuffled around and added an unknown offset $e_0$. 
We will take this $F$ to be the problem Hamiltonian of a QAA procedure:  
\beq
\hat{H}_p &=& e_0  \\\nonumber 
&+& \text{diag}[\pi(\underbrace{f_0,\ldots,f_0}_{m_0 \, \text{times}},\underbrace{f_1,\ldots,f_1}_{m_1 \, \text{times}},\ldots, \underbrace{f_K,\ldots,f_K}_{m_K \, \text{times}})] \,,
\eeq
where $\pi(\cdot)$ is an arbitrary permutation of its list of arguments.

The lack of structure exhibited by the problem Hamiltonian above does not seem to allow for any clever `tailoring' 
of QAA to the problem, as no information about the problem can be incorporated into the driver Hamiltonian
that would help in speeding up the computation. Therefore, a driver Hamiltonian that treats all computational-basis eigenvectors
in exactly the same way will presumably be an optimal choice.  
We thus take as driver Hamiltonian, the one-dimensional projection onto the equal-superposition state:
\beq
\hat{H}_d = -  E_0 | \phi \rangle \langle \phi |\,,
\eeq
where $E_0$ is a positive constant that provides a scale to the driver Hamiltonian and $|\phi \rangle$ is the fully-symmetric state 
\beq
| \phi \rangle = \frac1{\sqrt{N}} \sum_{i=1}^N | i \rangle \,.
\eeq
The ground-state of this driver Hamiltonian, i.e., the state that the system is prepared in, for the adiabatic evolution, is 
\beq
|\psi(t=0) \rangle = | \phi \rangle \,,
\eeq
whose energy is $-E_0$. 

We note here that both the scrambled problem Hamiltonian and the above driver Hamiltonian have been shown to pose severe limitations on
the efficiency of any QAA constructed using either of the two~\cite{farhi-2008-6}. We shall address this matter later on.
Here, however, the above choice of driver Hamiltonian has an attractive property, at least 
as far as scrambled problem Hamiltonians are concerned. Its symmetry makes it invariant under any permutation of the eigenstates of the computational (problem Hamiltonian) basis. Specifically, it is invariant under the (unknown) unitary transformation 
that `unscrambles', i.e., orders, the eigenvalues of the problem Hamiltonian. It thus follows that the Schr\"odinger equation of the QAA for this type of problems can be written down explicitly, and subsequently solved, at least in principle. This property of the system can then be used, as we show next,  to find a (not necessarily) adiabatic path $s(t)$ such that at the end 
of the algorithm, the final state of the system lies very close to (or in some cases, is precisely) the ground-state 
of the problem Hamiltonian. Once an optimal path is found, a measurement of the energy at the end of the run, 
will immediately reveal the minimum energy $e_0+f_0$ (and therefore also the value of the unknown offset $e_0$), 
along with the corresponding minimum-energy configuration. 

Explicitly, the Schr\"odinger equation for a QAA on a scrambled output problem, $i \frac{\rmd}{\rmd t} |\psi(t)\rangle =\hat{H} |\psi(t) \rangle$,
simplifies to the $(K+1)$ coupled first-order ordinary differential equations:
\beq \label{eq:schr}
i \dot{c}_j = -(1-s) \sum_{k=0}^K \eta_k c_k + s f_j c_j\,,
\eeq
where $j=0..n$. 
Here, $\eta_j=m_j/N$ are the ratios of the multiplicities of occurrences of the $j$-th eigenvalue $f_j$ to the total number of eigenvalues $N$. 
The various $c_j$ are the $(K+1)$ distinct wave-function amplitudes that correspond to the different values $f_j$. 
We have thus ended up with a compact set of equations which needs to be solved or analyzed in some meaningful way.


A few remarks are now in order: i) Note that the effective dimension of the problem is $K+1$, i.e., precisely the number of distinct $f_j$'s. 
ii) In terms of the different amplitudes $c_j$, the initial-state of the system corresponds to $c_j(t=0)=1$ for all $j$, where the normalization chosen here is $\sum_{j=0}^K \eta_j |c_j|^2=1$. Note that the evolving wave-function depends on $N$, the size of the system, only through the various ratios. iii) The addition of an unknown offset $e_0$ to the various $f_j$ only affects the (immaterial) global phase of the wave-function, and may therefore be completely removed from the equations.
 
In what follows, we utilize the ideas presented above, to obtain efficient analog algorithms for two exemplary problems 
for which no efficient classical solution is known. We first consider
the famous Deutsch-Josza problem~\cite{deutsch:92}, for which we find an 
efficient constant-runtime deterministic solution, that stands in contrast to the exponential complexity of the corresponding deterministic classical algorithm,
and is on par with the circuit-based quantum Deutsch-Josza algorithm. 
We next consider a variant of the random energy model already mentioned in the previous section, for which the minimum energy configuration is found quadratically faster than on a classical computer. 

\section{\label{sec:dj}An efficient deterministic algorithm for the Deutsch-Josza problem}

In the Deutsch-Josza problem, we are given a black box quantum computer (an oracle) that implements a boolean function of $n$ bits 
that is either constant, i.e., gives $0$ on all inputs or $1$ on all inputs, or balanced, i.e., returns $1$ for exactly half of the input states and $0$ for the other half. The task then is to determine with absolute certainty whether the function is constant or balanced with as few calls as possible to the oracle.

A classical deterministic algorithm 
would require $N/2+1$ evaluations of the function in the worst case (and two evaluations in the best case). 
The seminal result by Deutsch and Josza~\cite{deutsch:92} was the construction of a deterministic circuit-based quantum algorithm that requires only two calls to the oracle (later reduced to only one call by Cleve {\it et al.}~\cite{cleve:98}), thus providing the first example
of a quantum computation that is exponentially more efficient than the best corresponding classical algorithm.

In what follows, we show that one can use the principles discussed in the previous section to find an analog deterministic quantum computation
to solve the Deutsch-Josza problem with a runtime that is on par with the analogous circuit-based algorithm, i.e., a runtime that does not scale with input size, and so is exponentially faster than the corresponding deterministic classical algorithm.

The algorithm we propose here is very similar to adiabatic algorithms of the usual AQC paradigm but with one important distinction.
While with adiabatic processes the success  of the algorithm depends on the  state of the system
being close at all times to the instantaneous ground state of the evolving Hamiltonian, here  shall `relax' this condition.
We will show that by allowing the state of the system to `detach' itself at least to some extent from
the instantaneous ground state, one can in fact obtain a deterministic (i.e., zero probability of failure) non-adiabatic efficient algorithms,
for the Deutsch-Josza and certain other problems (see, e.g., Ref.~\cite{farhi:98}). 

\subsection{The adiabatic Deutsch-Josza equations}

In the context of `scrambled output' problems, the Deutsch-Josza oracle corresponds in the `constant' case to having $f_0=0$ or $f_0=1$ (with $m_0=2^n \equiv N$), and  $f_0=0$ and $f_1=1$ with multiplicity $m_0=m_1=N/2$ in the balanced case. The task here is to be able to 
distinguish between the two cases. In this problem, we shall assume, for simplicity, that the constant offset $e_0$ is zero (even though in principle it could take on any value). 

To find a deterministic efficient solution to the Deutsch-Josza problem, we construct a continuous-time algorithm that is optimized 
for the case where the input matrix $F$ is balanced, for reasons that will become clear shortly. 
In the balanced case, the problem Hamiltonian has two distinct values and the wave function $|\psi(t) \rangle$  has only two distinct components. 
It may therefore be written as 
\beq
|\psi(t) \rangle = c_0(t) | \psi_0\rangle  + c_1(t) | \psi_1 \rangle \,,
\eeq
where $|\psi_0 \rangle$ and $|\psi_1 \rangle$ are the equal superpositions of all solution (zero-energy) states and non-solution (energy one) states, respectively: 
\beq
|\psi_0 \rangle &\equiv& \sqrt{\frac{2}{N}} \sum_{m \in \mathcal{M}} | m\rangle \,,\\
|\psi_1 \rangle &\equiv& \sqrt{\frac{2}{N}} \sum_{m \notin \mathcal{M}} | m\rangle\,,
\eeq
where the $N/2$ states $m \in \mathcal{M}$ are eigenstates of the computational basis with zero eigenvalues (all other states have eigenvalue one).
The Schr\"odinger equation, Eq.~(\ref{eq:schr}), thus becomes the two coupled equations (here, we fix the energy scale of the driver Hamiltonian at $E_0=1$):
\beq
i \dot{c}_0 &=  -\frac1{2}(1-s) (c_0+c_1) & \\ 
i \dot{c}_1 &=  -\frac1{2}(1-s) (c_0+c_1) &+s c_1 \,,
\eeq
As previously mentioned, note that $N$, the size of the problem, does not appear in the equation.

The time-dependence of $c_0$ and $c_1$ is easily transformed 
into an $s$ dependence. 
The time-derivative transforms in the usual way: $\rmd/\rmd t = \left( \rmd s/\rmd t \right) \times \rmd/\rmd s$,
and the two coupled equations now become one, with the new complex-valued dependent variable $r(s)=c_1(s)/c_0(s)$:
\beq \label{eq:coupled}
2 i r'(s) = t'(s)\left[ -1+s +2 s r(s) +(1 - s) r(s)^2\right] \,,
\eeq
with the initial condition of $r(0)=1$ corresponding to the fully-symmetric initial state.
Here, the prime symbol $(')$ stands for differentiation with respect to the new independent variable $s$
and $t'(s) =1/\left(\rmd s/ \rmd t\right)$ encodes the non-adiabatic path.  

As a next step, we split the above equation into its real and imaginary parts, re-expressing the complex-valued $r(s)$ by 
real-valued components of magnitude and phase:
\beq
r(s)=\sqrt{ \frac{1-p(s)}{p(s)}}\exp^{i \phi(s)} \,.
\eeq
Here, $p(s)$ is the probability of the system to be in the symmetric zero-energy state $|\psi_0 \rangle$ at any given $s$,
and $\phi(s)$ is the phase, which is easily solved for, resulting in a single real-valued equation for the probability $p(s)$.

A deterministic algorithm whose end-state is precisely the equal superposition of the zero-energy states, $|\psi_0 \rangle$, may be obtained
if we satisfy the requirement that the final state of the system at $s=1$ 
will be precisely $p(1)=1$ (equivalently, $r(1)=0$). 
To make it so, we shall treat the Schr\"odinger equation, in this example Eq.~(\ref{eq:coupled}),
as an equation on the `path' $t'(s)$ rather than on the wave function itself [in our case, $p(s)$].
Put differently, we first choose the wave function ($r(s)$ here) or rather, the probability profile $p(s)$, such that the probability of obtaining the desired solution is exactly one at the end of the run,
and only then obtain the non-adiabatic path $t'(s)$ consistently with the Schr\"odinger equation of the system, Eq.~(\ref{eq:coupled}).

\begin{figure}
\begin{center}
\includegraphics[width=\figurewidth]{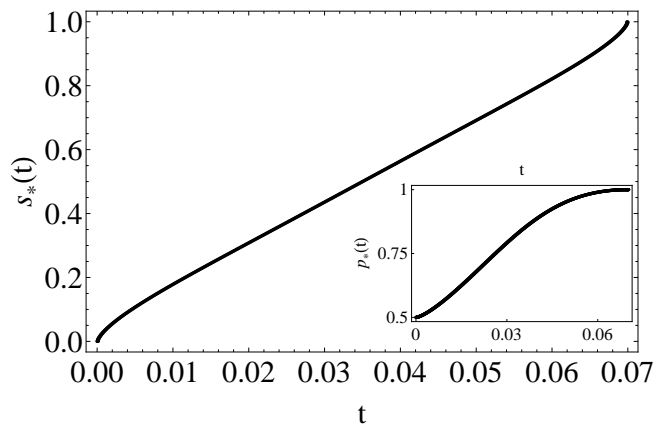}
\caption{The path $s_*(t)$ generated by the probability profile $p_*(s)$ given in Eq.~(\ref{eq:ps}).
The probability $p_*(s)$ is plotted in the inset.   
\label{fig:sOfT}}
\vspace{-.7cm}
\end{center}
\end{figure}

For the Deutsch-Josza problem, an explicit expression for the path $t'(s)$
as a function of the probability profile $p(s)$ may be obtained:  
\beq
t'(s) &=& \frac{p'(s)}{\sqrt{p(s)\left(1-p(s)\right)} \sqrt{1-q(s)^2}}\,,
\eeq
where 
\beq
q(s) = \frac{1-s(1+2 p(s) + 2(1-s) \int_0^s \frac{p('s)}{(1-s')^2} \rmd s'}{2 (1-s) \sqrt{p(s)\left(1-p(s)\right)}} \,.
\eeq
Presumably, there are infinitely many choices for $p(s)$ that yield a desired path. One however must make certain that $p(s)$ satisfies the boundary conditions, $p(0)=1/2$  (corresponding to $r(0)=1$), and $p(1)=1$ (corresponding to $r(1)=0$), and that the profile $t'(s)$ is well-defined everywhere and yields a finite running time for the algorithm, namely, that $\mathcal{T}=\int_0^1 t'(s) \rmd s < \infty$.
One example for such a probability profile is
\beq \label{eq:ps}
p_*(s)=\frac1{2}(1+6 s^2 - 8 s^3 + 3 s^4) \,,
\eeq
which immediately leads to 
\beq \label{eq:tp}
t_*'(s)=\frac{6 \sqrt{2}}{\sqrt{s (1-s)[4-9s(1-s)]}} \,.
\eeq
Figure~\ref{fig:sOfT} shows the ``adiabatic'' parameter $s_*(t)$ for the above example, obtained by integration and inversion of $t'_*(s)$ above. The inset shows $p_*(t)$, the generating probability profile. 

\begin{figure}
\begin{center}
\includegraphics[width=\figurewidth]{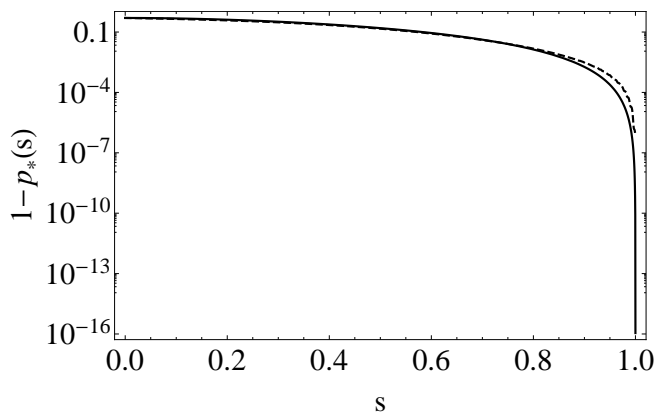}
\caption{Probability of being in the non-solution superposition,  $1-p_*(s)$, throughout the non-adiabatic evolution for the example discussed in the text (solid line)
compared against the probability profile of the adiabatic case (dashed line). While the latter probability reaches at the end of the process a small yet strictly nonzero
value at $s=1$, in the former case the probability of failure drops precisely to zero. 
\label{fig:det}}
\vspace{-.7cm}
\end{center}
\end{figure}

In Fig.~\ref{fig:det}, the probability of being in the non-solution superposition, namely, $1-p_*(s)$, is plotted as a function of $s$ on a logarithmic scale. As can be seen in the figure, this probability (solid line) drops sharply to zero as $s$ approaches one.
This is in contrast with 
the probability of failure in the adiabatic case (dashed line) generated by the adiabatic path $t'(s)=[\epsilon (1-2s(1-s))]^{-1}$ which does not vanish as $s=1$ (here, $\epsilon=0.001$).   

As mentioned above, there are presumably many paths $s(t)$ that produce, for a carefully determined runtime, the desired 
end probability of one. We note here, that even the simple choice of $s(t)=1/2$, while not conforming to the boundary 
conditions of $s(0)=1$ and $s(1)=1$, also produces the desired probability.  
The running time in this case, $\mathcal{T}=\sqrt{2} \pi \approx 4.443$, is however, substantially longer than the one found above.  

\subsection{The algorithm}

Now that it has been established that a deterministic algorithm to reach the zero-energy superposition 
for a balanced Deutsch-Josza problem Hamiltonian exists, we construct an efficient QAA for the Deutsch-Josza problem as follows. 
Within the quantum-analog Deutsch-Josza algorithm, one executes the QAA twice: Once with a problem Hamiltonian of $\hat{H}_p=F$ and 
a second time with $\hat{H}_p=1-F$. Note that if $F$ is balanced (constant) then $(1-F)$ is also balanced (constant). In both runs, the profile function to be used is one which guarantees the end-state to be the zero-energy superposition in the case where $F$ is balanced
(an example for such a path has been given above).

Now, if $F$ is constant, the problem Hamiltonian will have no effect on the state of the system in either run, regardless of the 
profile function chosen for the run. The state will remain in the initial state $|\phi \rangle$ throughout the evolution, except for an undetectable global phase. This is due to the constancy of the problem Hamiltonian.
A measurement of the $z$-magnetization at the end of the run will thus pick out a random bit configuration
that will have energy $0$ in one run and energy $1$ in the other. 

If on the other hand $F$ is balanced, the chosen path  
ensures us that at the end of each of the two runs, at $s=1$, the final state of the system will be the equal superposition of the zero-energy states with probability $1$. A measurement along the $z$-direction will thus produce in this case a zero-energy state in both runs. 

The resulting pair of energy measurements therefore distinguishes between a balanced function (both energy readouts are $0$) and a constant function (energy readouts of $0$ and $1$), deterministically.
Since $t'(s)$ is $N$-independent, the running time is $O(1) $, i.e., it does not scale with the size of the problem.
The above scheme thus provides a constant-runtime deterministic algorithm for solving the Deutsch-Josza problem. 

\section{\label{sec:rem}An efficient adiabatic solution to the random energy model}

In the following example, we consider a different model, namely, a variant of the random energy model~\cite{derrida:80,derrida:81},
previously considered under somewhat different settings in Refs.~\cite{farhi:11, jorg:10c} in the context of the quantum adiabatic algorithm. 
Here, we show that one can utilize the knowledge of the spectrum of the problem Hamiltonian, to obtain an optimal adiabatic
path that in turn yields a quantum adiabatic algorithm to find the minimum-energy configuration. The resulting algorithm
is found to be, as expected, quadratically-faster than any classical search algorithm. 

In this variant of the random energy model, the eigenvalues of the problem Hamiltonian are integers in the range 
$k \in \{{ 0..n \}}$ where each value $k$ appears precisely $\binom{n}{k}$ times. The $N=2^n$ eigenvalues are distributed randomly 
along the diagonal. As discussed earlier, one could further complicate the problem by considering adding an arbitrary constant offset to each of the $k$ values. 
Here too, one has to construct an efficient algorithm designed to find the minimum-energy bit configuration (i.e., the configuration that corresponds to $k=0$), and to determine how the running time of the algorithm scales with the dimension of the Hilbert space of the problem, $N=2^n$. Note that for the corresponding classical algorithm, the average running time would scale like $O(N)$, as the ground state has multiplicity one and there is no efficient searching routine to find it, due to the lack of structure of the problem. 

For what follows, we choose the energy scale of the driver Hamiltonian to be $E_0=n$ for computational convenience.
The Schr\"odinger equation, Eq.~(\ref{eq:schr}), becomes for this problem a set of only $n+1$ linear first-order differential equations,
corresponding to the $n+1$ distinct energies of the problem Hamiltonian. 
The characteristic polynomial of the reduced Hamiltonian, as a function of the adiabatic parameter $s$, is simplified to the equation: 
\beq
\frac{n(1-s)}{2^n}\sum_{k=0}^{n} \binom{n}{k}\frac{1}{k s-\lambda}=1\,.
\eeq
The solutions $\lambda$ to this equation are the eigenvalues of the Hamiltonian, and may be obtained 
analytically in the large $n$ limit. 

\begin{figure}
\begin{center}
\includegraphics[width=\figurewidth]{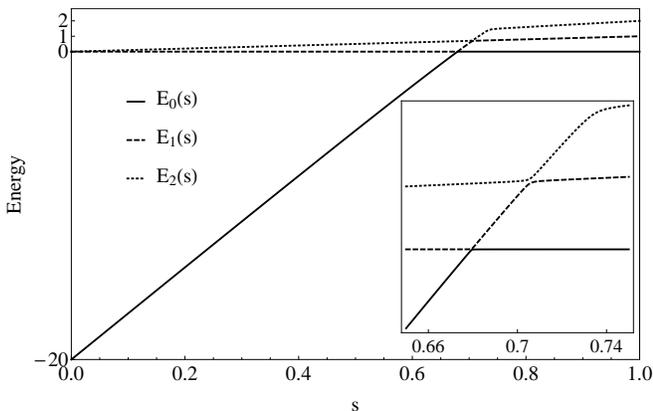}
\caption{The three lowest energy levels for the $n=20$ random energy model considered in the text. The inset is a blowup of the region 
of the of avoided crossings between the levels. \label{fig:e0e1e2}}
\vspace{-.7cm}
\end{center}
\end{figure}

Figure~\ref{fig:e0e1e2} shows the three lowest energy levels
of the system as a function of $s$ for $n=20$ (the energy levels are qualitatively similar for larger all $n$ values as well). The two avoided crossings between these three levels 
completely determine the behavior of the gap and the matrix element of the system for all $s$ values.
A simple perturbation analysis of the system reveals that one needs only to consider the projection
of the Hamiltonian into the subspace spanned by i) the ground state of the driver Hamiltonian, ii) the ground state of the problem Hamiltonian and iii) the equal superposition of the first $n$ excited states of the problem Hamiltonian (all of which have the same energy). Analysis of the three-level system in the large $n$ limit enables us to obtain rather easily the large-$n$ asymptotic behavior of the gap:
\beq \label{eq:g}
g(s)= \Bigg\{
\begin{tabular}{cccccc}
$n-\frac{3 n -1}{2}s$ & \phantom{A}  & $0$  & $<s \leq $  & $\frac{2n}{3n-1}$ \\
$\frac{3 n -1}{2}s-n$ & \phantom{A}  & $\frac{2n}{3n-1}$ & $<s \leq$ & $\frac{2n }{3n-3}$ \\
$s$ & \phantom{A} & $\frac{2 n}{3n-3}$ & $< s \leq $ & $1$ \\
\end{tabular}
\eeq
In the immediate vicinity of the minimum gap, we find that
\beq\label{eq:ming}
\min_s g(s)=\frac{2  n}{3} 2^{-n/2}\,,
\eeq
where the critical point is approaches $s_*=2/3$ for large $n$. 

While the matrix element $V_{01}(s)$ may also be computed using 
perturbation theory, it is easier to bound it from above by simpler considerations:
\beq
V_{01}(s) &=&| \langle 0 | \rmd H / \rmd s | 1\rangle| \\\nonumber
&=& \frac1{s} | \langle 0 | \hat{H}_d |1\rangle| =
\frac1{1-s}| \langle 0 | \hat{H}_p |1\rangle| \,.
\eeq
The above matrix elements for both the problem and driver Hamiltonian are easily bounded by their matrix norms to give: 
\beq
V_{01}(s) \leq \min_s \left[\frac{n}{s},\frac{n}{1-s}\right] =2 n\,.
\eeq
While the above bound is by no means tight (the actual tight bound is also linear in $n$ albeit with a smaller constant), the above bound will suffice for our purpose. 

A constant-rate adiabatic scheme, following Eq.~(\ref{eq:T}), will yield QAA running times that are $O(\max_s V_{01}(s) [  \epsilon \min_s g(s)^2]^{-1})=O(N/n \epsilon)$, and would therefore provide no real advantage over the performance of classical algorithms.

\begin{figure}
\begin{center}
\includegraphics[width=\figurewidth]{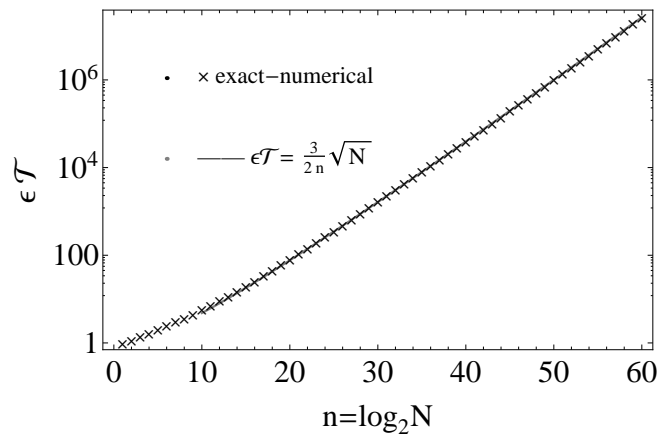}
\caption{The running time $\epsilon \mathcal{T}$ of the random energy model adiabatic algorithm as
a function of $n$, the number of input-bits, on a logarithmic scale. As the figure indicates, the running time scales like $\frac{\sqrt{N}}{n}$. \label{fig:runtime}}
\vspace{-.7cm}
\end{center}
\end{figure}

Nonetheless, our ability to calculate the gap of the system accurately, along with proper bounds on the matrix element,
allows for the construction of a locally-adiabatic path that would in turn yield significantly shorter QAA running times. 
The bound on the matrix element, combined with the gap $g(s)$ found in Eqs.~(\ref{eq:g}) and~(\ref{eq:ming}) ,
using the local adiabatic evolution condition, Eq.~(\ref{eq:lae}), allows us to construct the locally-adiabatic path 
which, upon integration, yields the following scaling on the running time in the large $n$ limit:
\beq \label{eq:T}
\mathcal{T} &=& \int_0^1 \rmd s \left( \frac{\rmd s}{\rmd t }\right)^{-1} \\\nonumber
&\sim& O(\max_s V_{01}(s) [  \epsilon \min_s g(s)]^{-1}) =O(\frac{\sqrt{N}}{n \epsilon}) \,.
\eeq
To verify that the large-scale analysis presented above is correct, 
 we plot in Fig.~\ref{fig:runtime} the actual running time as it was 
calculated by integrating the exact-numerical gaps and matrix elements  for up to $n=60$ (crosses, the $\epsilon$ pre-factor has been removed)  as a function of the number of input bits $n$, on a logarithmic scale.
The solid line in the figure corresponds to $\epsilon \mathcal{T}=\frac{3}{2 n} 2^{n/2-1}=\frac{3}{2n} \sqrt{N}$ which turns out to be 
the true large-$n$ behavior of the running time, and which provides a verification for the 
large $n$ scaling analysis presented above. 

We have thus shown how to construct a quantum adiabatic algorithm with which the random energy model can be solved
quadratically faster than any existing classical algorithm. 

We note here that the above analysis is in accord with theorems proved in Ref.~\cite{farhi-2008-6} by Farhi \textit{et al.} 
which have shown that the algorithmic efficiency of the QAA with either 
a one-dimensional projection as the driver Hamiltonian or an unstructured problem Hamiltonian (under certain conditions) is bounded by an $O(\sqrt{N})$ of the running time. Here, we have used both.

\section{\label{sec:conc}Summary and conclusions}
We have demonstrated how one can construct analog, continuous-time, quantum algorithms for a family of unstructured problems, namely, scrambled output. We have shown that for this type of problems, a compact set of dynamical equations may be written down explicitly, and subsequently solved or analyzed to enable finding optimized adiabatic or non-adiabatic paths,
that yield algorithms that are more efficient than corresponding classical algorithms.

We considered two specific examples. In the first, a simple prescription for solving the Deutsch-Josza problem was given. We have shown that the running time of the proposed analog algorithm does not scale with the size of the problem, i.e., that it is exponentially faster than the classical algorithm and on par with the circuit-based quantum algorithm result. 

In a second example, we provided an adiabatic solution
to the problem of finding the ground state of a random energy model, in which the eigenvalues are taken from a binomial distribution 
but are then shuffled around. In this example, the provided algorithm was found to be, as one might expect, quadratically faster than the corresponding classical one. 

It would be of interest to see how this family of problems may be further generalized, thereby expanding the scope 
of applicability of continuous-time quantum computing.
We hope that the
method presented here will open a way for
other adiabatic as well as deterministic non-adiabatic efficient algorithms that would further demonstrate the power and potential encompassed in analog continuous-time quantum computation.
Specifically, we hope that the above algorithm will help in further pinpointing the precise equivalence 
between circuit-based and adiabatic quantum computing. 

\begin{acknowledgments} 
We thank Peter Young, Eleanor Rieffel and David Gosset for useful comments and discussions. 
\end{acknowledgments}

\bibliography{refs}

\end{document}